\documentclass[11pt]{article}
\usepackage[margin=1in]{geometry}
\usepackage[utf8]{inputenc}
\usepackage{graphicx}
\usepackage{lineno}
\usepackage{url}
\usepackage{amssymb}
\usepackage[justification=centering]{caption}
\usepackage{threeparttable}

\begin{document}
\noindent
\textbf{\large{Relativistic Shapiro delay measurements of an extremely massive millisecond pulsar}}\\

\noindent
H.~T.~Cromartie$^{*1}$, E.~Fonseca$^{2}$, S.~M.~Ransom$^{3}$, P.~B.~Demorest$^{4}$, Z.~Arzoumanian$^{5}$, H.~Blumer$^{6,7}$, P.~R.~Brook$^{6,7}$, M.~E.~DeCesar$^{8}$, T.~Dolch$^{9}$, J.~A.~Ellis$^{10}$, R.~D.~Ferdman$^{11}$, E.~C.~Ferrara$^{12,13}$, N.~Garver-Daniels$^{6,7}$, P.~A.~Gentile$^{6,7}$, M.~L.~Jones$^{6,7}$, M.~T.~Lam$^{6,7}$, D.~R.~Lorimer$^{6,7}$, R.~S.~Lynch$^{14}$, M.~A.~McLaughlin$^{6,7}$, C.~Ng$^{15,16}$, D.~J.~Nice$^{8}$, T.~T.~Pennucci$^{17}$, R.~Spiewak$^{18}$, I.~H.~Stairs$^{15}$, K.~Stovall$^{4}$, J.~K.~Swiggum$^{19}$, \& W.~W.~Zhu$^{20}$

\begin{enumerate}
\setlength{\itemsep}{1pt}
\setlength{\parskip}{0pt}
\setlength{\parsep}{0pt}
\footnotesize{
\item \emph{Department of Astronomy, University of Virginia, 530 McCormick Rd., Charlottesville, VA 22903, USA}
\item \emph{Department of Physics, McGill University, 3600 University St., Montreal, QC H3A 2T8, Canada}
\item \emph{National Radio Astronomy Observatory, 520 Edgemont Rd., Charlottesville, VA 22903, USA}
\item \emph{National Radio Astronomy Observatory, 1003 Lopezville Rd., Socorro, NM 87801, USA}
\item \emph{X-ray Astrophysics Laboratory, Code 662, NASA Goddard Space Flight Center, Greenbelt, MD, 20771}
\item \emph{Department of Physics and Astronomy, West Virginia University, P.O. Box 6315, Morgantown, WV 26506, USA}
\item \emph{Center for Gravitational Waves and Cosmology, West Virginia University, Chestnut Ridge Research Building, Morgantown, WV 26505, USA}
\item \emph{Department of Physics, Lafayette College, Easton, PA 18042, USA}
\item \emph{Department of Physics, Hillsdale College, 33 E. College Street, Hillsdale, Michigan 49242, USA}
\item \emph{Infinia ML, 202 Rigsbee Avenue, Durham NC, 27701}
\item \emph{School of Chemistry, University of East Anglia, Norwich, NR4 7TJ, United Kingdom}
\item \emph{NASA Goddard Space Flight Center, Greenbelt, MD 20771, USA}
\item \emph{Department of Astronomy, University of Maryland, College Park, MD, USA}
\item \emph{Green Bank Observatory, P.O. Box 2, Green Bank, WV 24944, USA}
\item \emph{Department of Physics and Astronomy, University of British Columbia, 6224 Agricultural Road, Vancouver, BC V6T 1Z1, Canada}
\item \emph{Dunlap Institute, University of Toronto, 50 St. George St., Toronto, ON M5S 3H4, Canada}
\item \emph{Hungarian Academy of Sciences MTA-ELTE ``Extragalatic Astrophysics Research Group", Institute of Physics, E{\"o}tv{\"o}s Lor{\'a}nd University, P{\'a}zm{\'a}ny P. s. 1/A, 1117 Budapest, Hungary}
\item \emph{Centre for Astrophysics and Supercomputing, Swinburne University of Technology, P.O. Box 218, Hawthorn, Victoria 3122, Australia}
\item \emph{Center for Gravitation, Cosmology and Astrophysics, Department of Physics, University of Wisconsin-Milwaukee, P.O. Box 413, Milwaukee, WI 53201, USA}
\item \emph{CAS Key Laboratory of FAST, NAOC, Chinese Academy of Sciences, Beijing 100101, China}
}
\end{enumerate}


\textbf{Despite its importance to our understanding of physics at supranuclear densities, the equation of state (EoS) of matter deep within neutron stars remains poorly understood. Millisecond pulsars (MSPs) are among the most useful astrophysical objects in the Universe for testing fundamental physics, and place some of the most stringent constraints on this high-density EoS. Pulsar timing --- the process of accounting for every rotation of a pulsar over long time periods --- can precisely measure a wide variety of physical phenomena, including those that allow the measurement of the masses of the components of a pulsar binary system \cite{lor05}. One of these, called relativistic Shapiro delay \cite{sha64}, can yield precise masses for both an MSP and its companion; however, it is only easily observed in a small subset of high-precision, highly inclined (nearly edge-on) binary pulsar systems. By combining data from the North American Nanohertz Observatory for Gravitational Waves (NANOGrav) 12.5-year data set with recent orbital-phase-specific observations using the Green Bank Telescope, we have measured the mass of the MSP J0740+6620 to be $\mathbf{2.14^{+0.10}_{-0.09}}$ solar masses (68.3\% credibility interval; 95.4\% credibility interval is $\mathbf{2.14^{+0.20}_{-0.18}}$ solar masses). It is highly likely to be the most massive neutron star yet observed, and serves as a strong constraint on the neutron star interior EoS.}

Relativistic Shapiro delay, which is observable when a pulsar passes behind its stellar companion during orbital conjunction, manifests as a small delay in pulse arrival times induced by the curvature of spacetime in the vicinity of the companion star. For a highly inclined MSP-white dwarf binary, the full delay is of order $\sim$10\,$\mu$s. The relativistic effect is characterized by two parameters, ``shape" and ``range." In general relativity, shape ($s$) is the sine of the angle of inclination of the binary orbit ($i$), while range ($r$) is proportional to the mass of the companion, $m_{\rm c}$. When combined with the Keplerian mass function, measurements of $r$ and $s$ also constrain the pulsar mass ($m_{\rm p}$; \cite{fre10} provides a detailed overview and an alternate parameterization).

Precise neutron star mass measurements are an effective way to constrain the EoS of the ultra-dense matter in neutron star interiors. Although radio pulsar timing cannot directly determine neutron star radii, the existence of pulsars with masses exceeding the maximum mass allowed by a given model can straightforwardly rule out that EoS. 

In 2010, Demorest et al.~reported the discovery of a 2-solar-mass MSP, J1614$-$2230 \cite{dem10} (though the originally reported mass was $1.97 \pm 0.04$\,M$_{\odot}$, continued timing has led to a more precise mass measurement of $1.928 \pm 0.017$\,M$_{\odot}$; Fonseca et al.~2016 \cite{fon16}). This Shapiro-delay-enabled measurement disproved the plausibility of some hyperon, boson, and free quark models in nuclear-density environments. In 2013, Antoniadis et al.~used optical techniques in combination with pulsar timing to yield a mass measurement of $2.01 \pm 0.04$\,M$_{\odot}$ for the pulsar J0348+0432 \cite{ant13}. These two observational results (along with others; see \cite{fri11}) encouraged a reconsideration of the canonical 1.4-M$_{\odot}$ neutron star. Gravitational wave astrophysics has also begun to provide EoS constraints; for example, the Laser Interferometer Gravitational-Wave Observatory (LIGO) detection of a double neutron star merger constrains permissible equations of state, suggesting that the upper limit on neutron star mass is 2.17\,M$_{\odot}$ (90\% credibility; \cite{mar17}). Though the existence of extremely massive ($>\,2.4$\,M$_{\odot}$) neutron stars has been suggested through optical spectroscopic and photometric observations (e.g.~\cite{lin18}), radio timing can provide much more precise constraints on the existence of $\gtrsim$\,2\,M$_{\odot}$ neutron stars.

NANOGrav employs pulsar timing for an important general relativistic application: the detection of low-frequency gravitational waves primarily from supermassive black hole binaries. The collaboration's observing program consists of high-cadence, multi-frequency radio observations of $\sim$75 MSPs using the Green Bank and Arecibo telescopes (GBT and AO; see \cite{arz18} and the upcoming 12.5-year data release). Additionally, NANOGrav has begun using the Karl G.~Jansky Very Large Array as the third observatory in its pulsar timing program. Using the Green Bank Telescope, NANOGrav regularly observes J1614$-$2230 and another high-mass radio MSP, J0740+6620.

PSR J0740+6620 (period = 2.89\,ms) was discovered in the Green Bank Northern Celestial Cap 350-MHz survey (GBNCC) in 2012 \cite{sto14}. It is in a nearly circular (eccentricity = $5 \times 10^{-6}$), 4.77-day orbit (Lynch et al.~presented a recent GBNCC timing solution in 2018 \cite{lyn18}). Recent optical and near-infrared observations by Beronya et al.~(2019) revealed that its companion is likely the coolest white dwarf ever measured in orbit with an MSP \cite{ber19}.

Here we present timing observations of the pulsar with the GBT taken between 2014 and 2019. We observed the pulsar regularly throughout this period as part of the NANOGrav timing program \cite{arz18}. This section of our data set includes $\sim$70 epochs (occurring approximately monthly and at random orbital phases) during which the pulsar was observed at both 1.4 GHz and 820 MHz for $\sim$20 minutes each. We were awarded additional time for two concentrated campaigns over superior conjunction (i.e.~when the pulsar is behind its companion star), as probing the minima and maxima of the Shapiro delay signal is the best way to improve sensitivity to it (see the absorbed or ``detectable" signal in the second panel of  Figure 1). 

After the second concentrated campaign consisting of two five-hour observations at orbital phases 0.15 and 0.25 (GBT 18B--372), the timing analysis (see details in Methods) yielded a pulsar mass of 2.14$^{+0.10}_{-0.09}$\,M$_{\odot}$ at 68.3\% credibility. The Methods section describes our rationale for choosing these two orbital phases, as well as the progression of mass measurements and precisions as more observations were added. Our final fits with and without Shapiro delay as a function of orbital phase are presented in Figure 1, and the top panel of Figure 2 shows timing residuals spanning the entire data set. 
Although our measured relative uncertainty is higher than, for example, the original relative error reported by Demorest et al.~for J1614$-$2230 (5\% vs.~2\%), J0740+6620 is a remarkably high-mass MSP. This measurement will help constrain high-density nuclear physics, as there are very few examples of $\gtrsim\,2$\,M$_{\odot}$ neutron stars. PSR J0740+6620 is 98\% and 90\% likely to be more massive than J1614$-$2230 and J0348+0432, respectively, and is therefore likely to be the most massive well measured neutron star to date. 

Taken together, these three massive MSPs serve as a strong validation of the existence of high-mass neutron stars. Due to the asymptotic nature of the relationship between maximum neutron star mass and nearly all EoS, even small increases in the measured mass of the most massive neutron stars force a reconsideration of the fundamental physics at play in their interiors (for example, see Figure 3 in \cite{wat15}). Non-nucleonic solutions to the EoS problem, such as quark matter, hyperons, or Bose-Einstein meson condensates, yield softer equations of state (i.e.~relatively compressible matter); however, more massive neutron stars necessitate stiffer EoS, which allow for higher maximum masses (see \cite{oze16} for a review). The measurement of a 2.14-solar-mass neutron star is therefore in extreme tension with these non-nucleonic proposals, and underlines the necessity of untangling existing theoretical paradoxes. The most prominent of these may be the hyperon problem, which proposes that although the extreme densities inside neutron stars would favor the conversion of nucleons to hyperons, the presence of hyperons softens the EoS and excludes the possibility of high-mass neutron stars (for example, \cite{bed15}). In addition, the mass measurement of J0740+6620 may have implications for the nature of neutron star mergers as detected by LIGO. Because several neutron stars with masses close to or greater than $\sim$2\,M$_{\odot}$ are now known, it may be the case that more mass-asymmetric neutron star mergers will occur than previously supposed. 

Constraining the mass of J0740+6620 carries additional astrophysical benefits. Recent evidence from Antoniadis et al.~(2016) \cite{ant16} suggests that the distribution of MSP masses may be bimodal, implying that many more neutron stars with masses greater than $\sim$1.6\,M$_{\odot}$ may exist than previously supposed (see also \cite{oze16}). Not only is it becoming clear that high-mass neutron stars make up a significant portion of the population, but their existence also carries substantial implications for our understanding of MSP binary evolution. Because many fully recycled pulsars have been measured to have masses less than or equal to 1.4\,M$_{\odot}$, we know that recycling can be accomplished with only a small amount of mass transfer. We must therefore consider the possibility that some MSPs are not formed near the Chandrasekhar mass and increase to high masses through accretion; rather, they are born massive in the first place (see \cite{tau11} and \cite{cog17} for earlier evidence of this phenomenon).

There exists a well known relationship between the mass of a pulsar's white dwarf companion and the binary system's orbital period (\cite{rap95}, \cite{tau99}). For our measured orbital period of $\sim$4.77 days, the predicted white dwarf companion masses (from equations 20 and 21 in Tauris \& Savonije 1999 \cite{tau99}) are $\sim$0.24\,M$_{\odot}$ for a mid-metallicity (Pop I+II) donor star and $\sim$0.25\,M$_{\odot}$ for a low-metallicity, Pop II star. Our measured mass of J0740+6620's helium white dwarf companion is $0.260^{+0.008}_{-0.007}$\,M$_{\odot}$ (at 68.3\% credibility). Given the stated uncertainties in convective mixing length, this discrepancy of 5$-$10\% is not an indication that J0740+6620 is an exception to the orbital period vs.~white dwarf mass relationship; however, it may indicate that this system was born in a relatively low-metallicity environment. There exist at least three other examples of MSP-helium white dwarf binaries with minimum companion masses greater than the Pop II masses predicted by Tauris \& Savonije (J1125$-$6014, J1903$-$7051, and J1933$-$6211). Lastly, if J0740+6620 is measured to be at the high end of our mass credibility interval, it may provide evidence that the creation of a stable, high-mass neutron star is possible through the merger of two low-mass neutron stars (in a LIGO-like gravitational wave event). 

Though it will require significant additional observing time to improve upon our J0740+6620 measurement, high-cadence monitoring of the pulsar is a promising strategy. Daily observations with the Canadian Hydrogen Intensity Mapping Experiment (CHIME; see \cite{ng18}) telescope, in conjunction with the present data set, have the potential to determine the mass of J0740+6620 with 2-3\% precision within a year. Additionally, the Neutron Star Interior Composition Explorer (NICER) is observing J0740+6620 at X-ray wavelengths (\url{https://heasarc.gsfc.nasa.gov/docs/nicer/science_team_investigations}). Modeling the thermal pulse profile of this MSP at X-ray energies will aid in constraining the mass and radius of J0740+6620. Continued collaboration with multifrequency observing programs will guarantee the steady improvement of this pulsar mass measurement in the long term.

\subsection*{Methods}

\textbf{\emph{Green Bank Telescope Observations.}} Both NANOGrav and targeted observations were conducted using the Green bank Ultimate Pulsar Processing Instrument (GUPPI, \cite{dup08}). Observations at 1500 MHz were acquired with 800 MHz of bandwidth split into 512 frequency channels (which were summed to 64 channels before analysis), sampling every 0.64\,$\mu$s. At an observing frequency of 820 MHz, 200 MHz of bandwidth over 128 channels was acquired with an identical sampling rate (and later also summed to 64 channels). These dual-polarization observations at both frequencies were coherently dedispersed at the known DM of 15.0\,pc\,cm$^{-3}$. Data were processed using NANOGrav pipelines for consistency with the existing four-year-long NANOGrav J0740+6620 data set (see \cite{arz15} for a thorough description of NANOGrav observing procedures, and \cite{dem18} for a description of NANOGrav's main data processing pipeline, \texttt{nanopipe}).\\

\textbf{\emph{Generation of TOAs and the Timing Model.}} The measurement and modeling of pulse times of arrival (TOAs) closely mirrors the procedure described by Arzoumanian et al.~2018 \cite{arz18}. We provide a summary of the analysis procedure in this section.

During offline processing, total-intensity profile data were integrated over $\sim$20--30 minute intervals to yield one or two TOAs per downsampled frequency interval for a normal NANOGrav observation, and $\sim$10 minutes for the long scans near or during conjunction. We extracted TOAs from each of the 64 integrated channels over the entire observing bandwidth through cross correlation between the data and a smoothed profile template using the software package \texttt{PSRCHIVE} (source code in \cite{van11}; see \url{http://psrchive.sourceforge.net}).

We used standard pulsar-timing analysis tools, namely  \texttt{TEMPO} (\url{http://tempo.sourceforge.net}) and \texttt{TEMPO2} (source code in \cite{hob12}; see \url{https://www.atnf.csiro.au/research/pulsar/tempo2}) for modeling TOA variation in terms of many physical mechanisms. \texttt{TEMPO} and \texttt{TEMPO2}, while not fully independent timing packages, yield consistent results. For J0740+6620, fitted parameters include: celestial (ecliptic) coordinates; proper motion; spin frequency and its first derivative; and binary orbital parameters (see Table 1 which lists best-fit values for these parameters as determined with \texttt{TEMPO}). 

We used the DE436 (\url{https://naif.jpl.nasa.gov/pub/naif/JUNO/kernels/spk/de436s.bsp.lbl}) Solar System ephemeris, maintained by the NASA Jet Propulsion Laboratory, for correction to the barycentric reference frame. The time standard used was BIPM2017. The overall RMS timing residual value for the timing model presented in this work is  1.5\,$\mu$s. The $\chi^2$ of our fit is 7314.35 with 7334 degrees of freedom, yielding a reduced-$\chi^2$ value of 0.997; note that the noise modeling (see Assessment of Timing Noise) will always yield a $\chi^2$ of $\sim$1.

We employed the ELL1 binary timing model \cite{lan01} in describing the nearly-circular orbital dynamics of the J0740+6620 system. Parameters of the ELL1 binary model consist of the projected semi-major axis, orbital period, epoch of passage through the ascending orbital node, and two ``Laplace-Lagrange parameters" ($\epsilon_1$ and $\epsilon_2$; the orbital eccentricity multiplied by the sine and cosine of periastron longitude, respectively; \cite{lan01}) that quantify departures from perfectly circular orbits.\\

\textbf{\emph{Assessment of Timing Noise.}} MSP rotation often exhibits a limit in achievable precision due to the presence of stochastic processes that act as noise to timing measurements. Examples of timing noise include systematic errors from cross-correlation template matching and ``spin noise" due to irregular rotation of the neutron star. We use a noise model similar to those developed in the NANOGrav 9-year and 11-year data releases in order to quantify these noise terms in the J0740+6620 data set.

The noise model consists of white-noise components that combine to form additive Gaussian noise. For each of the two frontend receivers used in this work, we use three parameters to describe the white-noise contribution to timing noise: a scaling factor applied to all raw TOA uncertainties (``EFAC"); a term added in quadrature to the TOA uncertainties (``EQUAD"); and a noise term that quantifies TOA correlations purely across observing frequency (``ECORR").

We used the \texttt{Enterprise} (\url{https://enterprise.readthedocs.io/en/latest}) modeling suite for estimation of the white components of the noise model using a Markov chain Monte Carlo (MCMC)-based algorithm. Enterprise uses the \texttt{TEMPO(2)} fit as the maximum-likelihood fit for the timing parameters and the basis of the fit for the red noise parameters, should they be found to be significant. In our \texttt{TEMPO(2)} fits, we include an EFAC of 1.036 for L-band (1500-MHz) TOAs and 1.013 for 820-MHz TOAs. EQUAD for L-band is 0.00610\,$\mu$s, and 0.18310\,$\mu$s for 820 MHz. ECORR values for L-band and 820-MHz TOAs are 0.00511\,$\mu$s and 0.00871\,$\mu$s, respectively. Bayesian model selection via an \texttt{Enterprise} MCMC run disfavors the inclusion of red noise; therefore, the noise model includes only white noise components.\\

\textbf{\emph{Dispersion Measure Modeling.}} The complexity of modeling DM variations arising from a dynamic interstellar medium has been discussed at length in previous works (see, for example, Lam et al.~2016 and Jones et al.~2017 \cite{lam15,jon17}). We have adopted the standard NANOGrav piecewise-constant model for DM trends wherein each epoch of data is fit with a constant ``DMX” value; in other words, each of these parameters is a deviation from some nominal DM and is fixed over a single epoch. The observation that J0740+6620's DM behavior is somewhat smooth over the duration of our data set (see Figure 2) led us to attempt alternatively modeling the entire data set by fitting for only the first and second derivatives of DM. In theory, this approach could be advantageous given the ability of DMX to absorb Shapiro delay signals (thanks to the similar duration of conjunction and a DMX epoch). While this strategy does reduce the formal parameter uncertainties from the fit, both an F-test and an Akaike information criterion test strongly favor the DMX model over the quadratic DM fit. This indicates the DM variation is not fully characterized by a quadratic model, and parameter values (including pulsar mass) derived from this model are likely to have systematic biases not reflected in their formal uncertainties.\\
\\
\textbf{\emph{Simulations.}} Analysis of the NANOGrav 12.5-year data set without supplemental data yielded $m_{\rm p}$ = $2.00 \pm 0.20\,$M$_{\odot}$. After the initial 6-hour supplemental observation, we measured the mass of J0740+6620 to be $2.18 \pm 0.15\,$M$_{\odot}$. We conducted simulations of future observations both to predict the constraining power of a concentrated Director's Discretionary Time campaign as well as to determine how our mass measurement may improve with additional observations going forward. For these simulations, we first generated an arbitrary array of TOAs that mirror the desired observing cadence, starting date, etc. The TOAs were then fit (with pulsar timing software such as \texttt{TEMPO} or \texttt{PINT}; \url{https://github.com/nanograv/PINT}) using the known parameters for J0740+6620. Residuals from this fit were then subtracted from the original TOAs to create ``perfect" TOAs, to which stochastic noise was then added. 
Two notable types of simulations were conducted. The first was an estimation of the improvement in our measurement of $m_{\rm p}$ given random orbital sampling (the ``NANOGrav-only observation" scenario); this solidified our conclusion that the concentrated GBT campaigns were necessary. The second served to optimize our observing strategy during a targeted orbital phase campaign by trying various permutations of orbital phase, number of observing sessions, and observing session lengths. The results of this simulation informed our GBT Director's Discretionary Time request for five hours over conjunction and five hours in one of the Shapiro ``troughs" (we were awarded time in the first trough --- around orbital phase 0.15 --- in addition to conjunction). In order to ensure that obtaining data in this asymmetric fashion would not bias our mass measurement, we ran 10,000 simulations of a five-hour conjunction observation plus five hours in either the first or second Shapiro trough. The averages of the 10,000 mass measurements obtained from each of these troughs were consistent within 1\%, implying that our orbital sampling is not biasing our results (as one would expect, given that the Shapiro delay response curve is symmetric about superior conjunction).\\
\\
\textbf{\emph{Data Availability.}} PSR J0740+6620 TOAs from both the 12.5-year data set and from the two supplemental Green Bank Telescope observations will be available at \url{https://data.nanograv.org} upon publication of this manuscript.
\\
\\
\textbf{\emph{Code Availability.}} All code mentioned in this work is open source and available at the links provided in the manuscript. 



\subsection*{Correspondence}
Correspondence and requests for materials should be addressed to H.~Thankful Cromartie (\url{thankful@virginia.edu}).

\subsection*{Acknowledgements}
The NANOGrav Project receives support from NSF Physics Frontiers Center award number 1430284. Pulsar research at UBC is supported by an NSERC Discovery Grant and by the Canadian Institute for Advanced Research (CIFAR). The National Radio Astronomy Observatory and the Green Bank Observatory are facilities of the National Science Foundation operated under cooperative agreement by Associated Universities, Inc. S.M.R is a CIFAR Senior Fellow.  W.W.Z.~is supported by the CAS Pioneer Hundred Talents Program, the Strategic Priority Research Program of the Chinese Academy of Sciences, grant No.~XDB23000000, and the National Natural Science Foundation of China grant No.~11690024, 11743002, 11873067. Supplementary Green Bank conjunction-phase observing project codes were 18B-289 and 18B-372 (DDT).

\subsection*{Author Contributions}
The creation of the NANOGrav 12.5-year data set was made possible through extensive observations and pulsar timing activities conducted by the authorship list in its entirety. H.T.C.~was responsible for the NANOGrav-adjacent concentrated observing campaigns and the majority of this manuscript's contents. H.T.C., E.F., S.M.R., and P.B.D.~were responsible for the extended J0740+6620 data analysis (the merging of NANOGrav and conjunction-phase observations) and modeling effort. E.F.~was responsible for much of the initial work on J0740+6620 that informed the supplementary observing proposals, and for the development of the gridding code that yielded both the mass and inclination credibility intervals and Figure 3. 
\subsection*{Competing Interests}
The authors of this letter declare no competing interests.

\newpage

\centering
\begin{table}[ht]
\centering
\caption{PSR J0740+6620 Best-Fit Parameters}
\begin{threeparttable}
\begin{tabular}{ll}
\hline
\hline
Pulsar name\dotfill & J0740+6620\\
Dates of Observations (MJD)\dotfill & 56640 -- 58462\\
Number of TOAs\dotfill & 7419\\
\hline
\multicolumn{2}{c}{Measured Quantities} \\ 
\hline
Ecliptic longitude, $\lambda$ (degrees)\dotfill & 103.75913607(1) \\
Ecliptic latitude, $\beta$ (degrees)\dotfill & 44.10248468(2) \\
Epoch of position \& period (MJD)\dotfill & 57551.0 \\
Proper motion in ecliptic longitude (mas yr$^{-1}$)\dotfill & $-$2.75(3) \\
Proper motion in ecliptic latitude (mas yr$^{-1}$)\dotfill & $-$32.43(4) \\
Parallax (mas)\dotfill & 0.5(3) \\
Spin frequency, $\nu$ (Hz)\dotfill & 346.5319964932129(6) \\
Spin frequency derivative, $\dot{\nu}$ (s$^{-2}$)\dotfill & $-$1.46389(2)$\times$10$^{-15}$ \\
Dispersion measure, DM (pc\,cm$^{-3}$)* \dotfill & 14.961787 \\
Profile frequency dependency parameter, FD1\dotfill & $-$1.17(4)$\times$10$^{-5}$ \\
Binary model\dotfill & ELL1\\
Projected semi-major axis of orbit, $x$ (lt-s)\dotfill & 3.9775561(2) \\
Binary orbital period, $P_{\rm{b}}$ (days)\dotfill & 4.7669446191(1) \\
Epoch of ascending node, TASC (MJD)\dotfill & 57552.08324415(2) \\
EPS1 (first Laplace-Lagrange parameter), $e\sin{\omega}$\dotfill & $-$5.70(4)$\times$10$^{-6}$ \\
EPS2 (second Laplace-Lagrange parameter), $e\cos{\omega}$\dotfill & $-$1.89(3)$\times$10$^{-6}$ \\
Sine of inclination angle $i$\dotfill & 0.9990(2) \\
Companion mass, $m_{\rm{c}}$ (M$_{\odot}$)\dotfill & 0.258(8) \\
\hline
\multicolumn{2}{c}{Derived Parameters} \\
\hline
Orbital eccentricity, $e$\dotfill & 5.10(3)$\times$10$^{-6}$ \\
Longitude of periastron, $\omega$ (degrees)\dotfill & 244.4(3) \\
Epoch of periastron, $T0$ (MJD) \dotfill & 57550.543(5)\\ Binary mass function (M$_{\odot}$)\dotfill & 0.0029733870(4) \\
Pulsar mass (68.3\% credibility interval, M$_{\odot}$)\dotfill & 2.14$^{+0.10}_{-0.09}$\\
Pulsar mass (95.4\% credibility interval, M$_{\odot}$)\dotfill & 2.14$^{+0.20}_{-0.18}$\\
Companion mass (68.3\% credibility interval, M$_{\odot}$)\dotfill & 0.260$^{+0.008}_{-0.007}$\\
Companion mass (95.4\% credibility interval, M$_{\odot}$)\dotfill & 0.260$^{+0.016}_{-0.014}$\\
Inclination angle (68.3\% credibility interval, degrees)\dotfill & 87.38$^{+0.20}_{-0.22}$\\
Inclination angle (95.4\% credibility interval, degrees)\dotfill & 87.38$^{+0.39}_{-0.45}$\\
\hline
\end{tabular}
\begin{tablenotes}
\item[]
\item[]\footnotesize{*Because this DM is an unfitted reference value, no error is reported. Values of DMX for each of the $\sim$70 epochs are available upon request.}
\end{tablenotes}
\end{threeparttable}
\label{params}
\end{table}

\newpage

\begin{figure}
    \centering
    \includegraphics[width=0.85\linewidth]{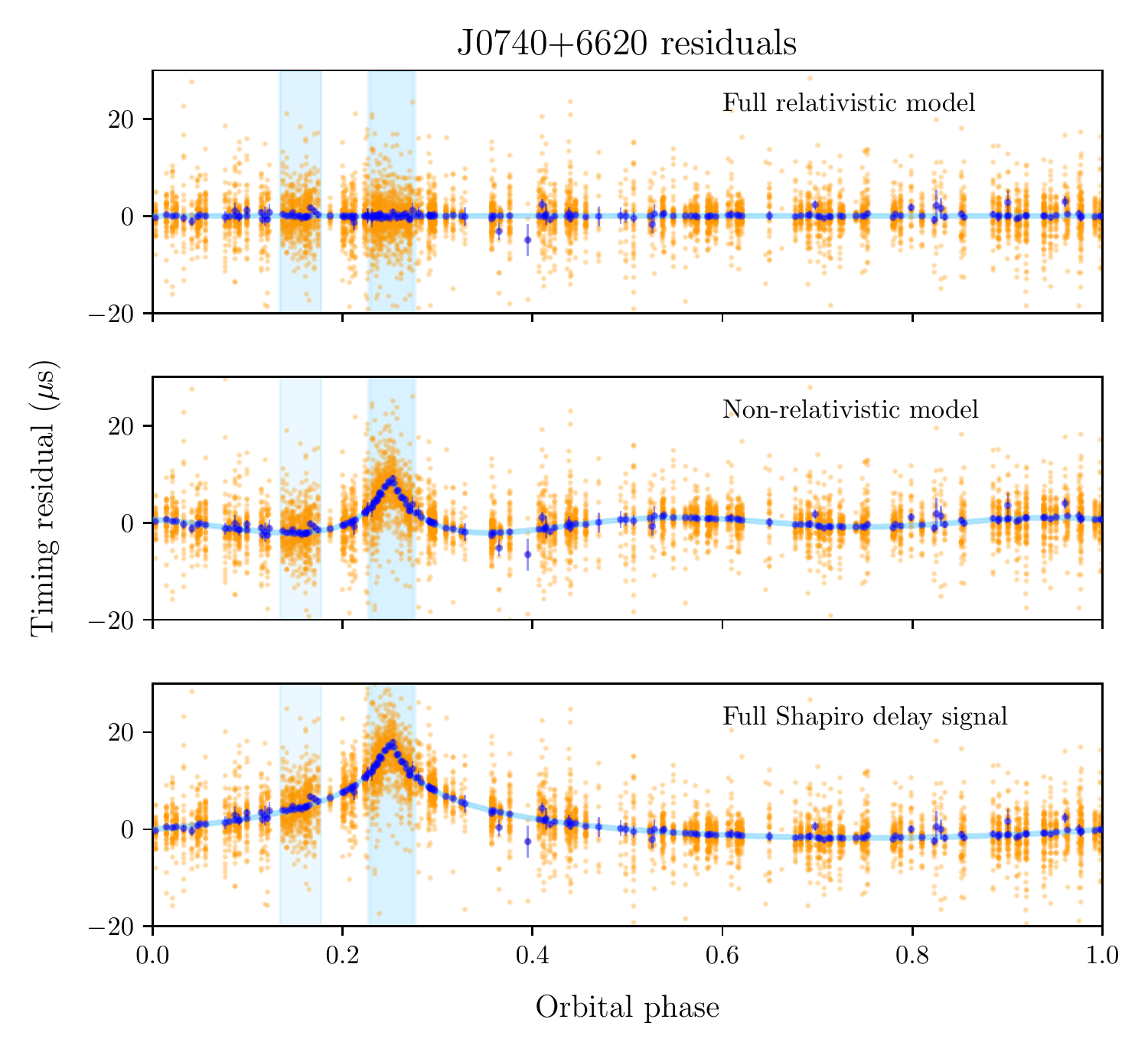}
    \caption{Timing residuals from all observations of J0740+6620 as a function of orbital phase, with superior conjunction at orbital phase = 0.25. Orange points are multi-frequency timing residuals, while dark blue points are averages of each group (i.e.~timing epoch) of these points with 1-$\sigma$ error bars. Averages were taken over a minimum of four data points to avoid showing misleading residuals from faint observations. Blue boxes indicate the orbital phases over which each of the three supplemental observations were taken (the box over conjunction is slightly darker because we made two superior conjunction observations). The top panel shows the full fit (including Shapiro delay parameters and all dispersion measure parameters --- i.e.~the full timing solution). The middle panel is the best fit with the measurable Shapiro delay signal added; this is the signal to which we are actually sensitive. The bottom panel is the ``full” Shapiro delay signal. Both the second and third panels are calculated based on the orbital and system parameters determined from the full fit. The lighter blue line in the middle and bottom panels represents the theoretical measurable and full Shapiro delay, respectively (and marks a 0-$\mu$s residual in the top panel). The width of the line in each panel is equal to the root mean squared error of the averaged points.}
    \label{fig:resids}
\end{figure}

\begin{figure}
    \centering
    \includegraphics[width=\linewidth]{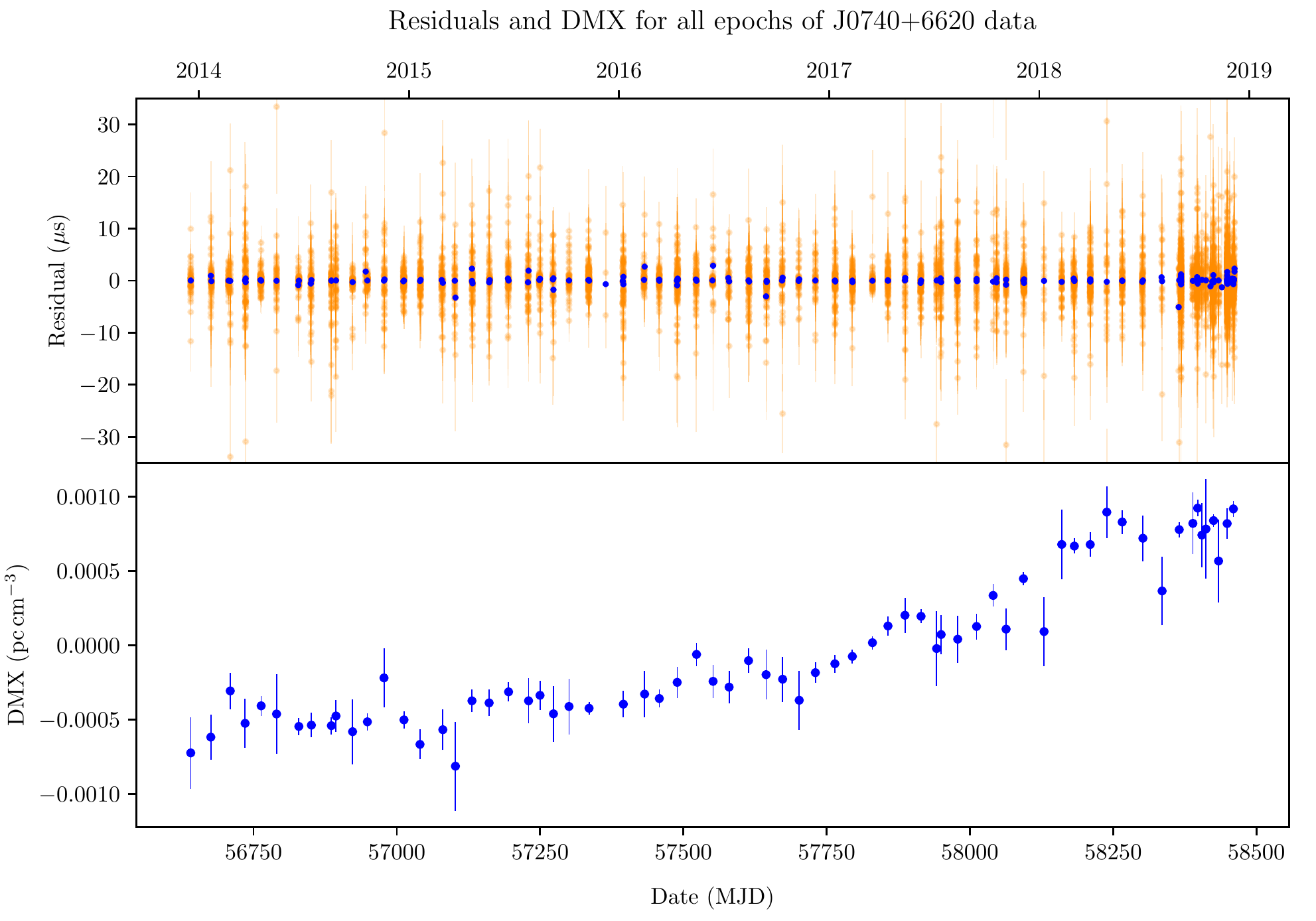}
    \caption{Timing residuals and DMX for all epochs of J0740+6620 data. \emph{Top panel:} Timing residuals from all epochs of J0740+6620 data, including both NANOGrav and superior conjunction-specific observations at all frequencies, are shown in orange (with 1-$\sigma$ error bars). The superimposed blue points represent an average over each epoch (RMS = 1.5\,${\mu}$s; note that some days have two separately calculated averages from dual-frequency data). \emph{Bottom panel:} Blue points indicate DMX values calculated for each epoch of data with 1-$\sigma$ error bars. The DMX trend is somewhat simple (i.e., roughly quadratic); however, linear modeling is strongly disfavored. A single averaged epoch (one dark blue point) was removed from these plots, as its error bar was $\sim$8\,$\mu$s due to a faint detection from which only one TOA could be extracted.}
    \label{fig:toa}
\end{figure}

\begin{figure}
    \centering
    \includegraphics[width=\linewidth]{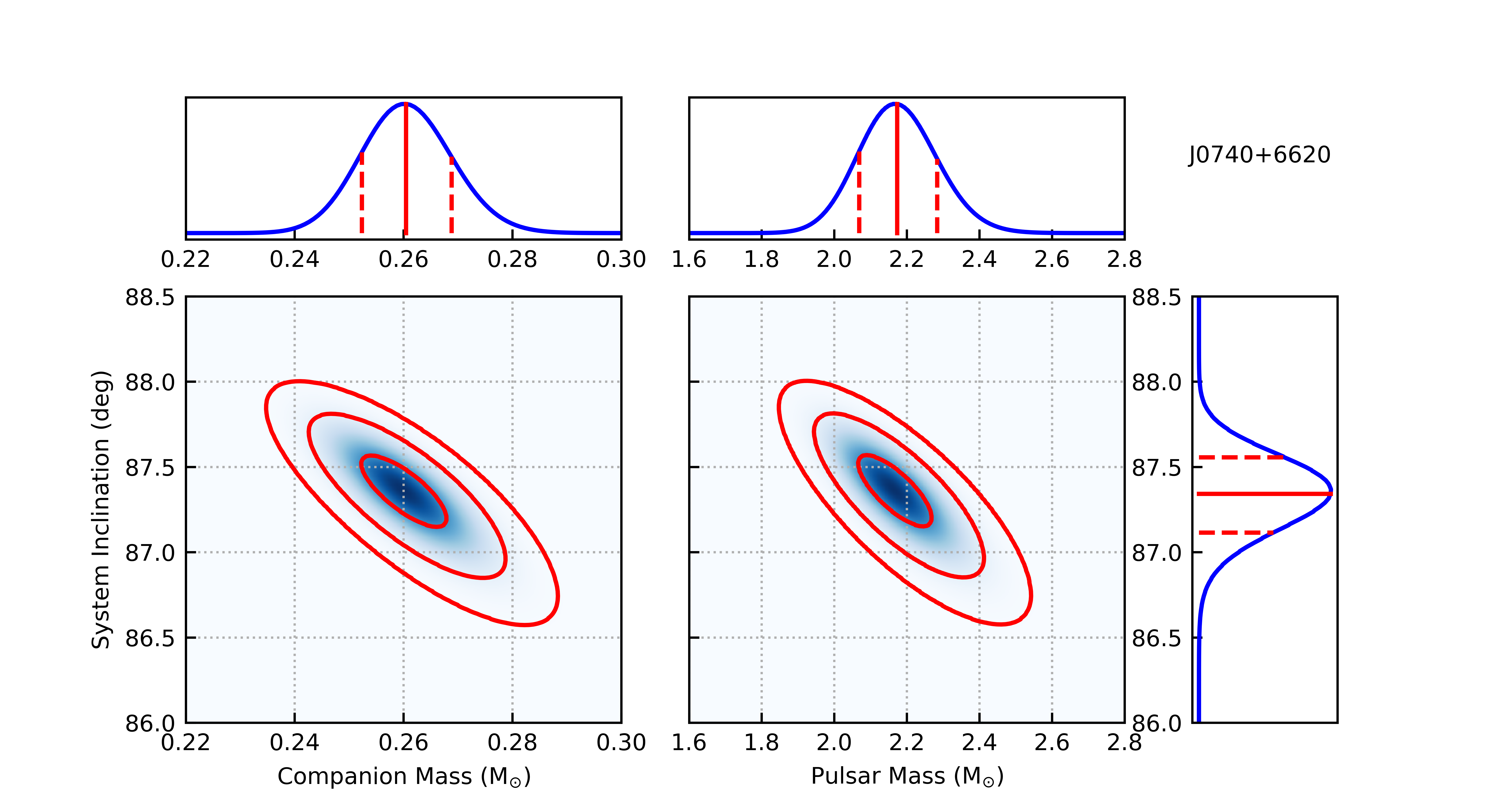}
    \caption{Map of fitted $\chi^2$ distributions and corresponding probability density functions for $m_{\rm p}$, $m_{\rm c}$, and $i$. The left-hand heat map was generated by computing $\chi^2$ values for different combinations of $m_{\rm c}$ and $i$; the right-hand heat map was calculated by translating the $m_{\rm c}-i$ probability density function to the $m_{\rm p}-i$ phase space using the binary mass function. Darker blue regions correspond to lower $\chi^2$ values. The three red circles correspond to 1, 2, and 3-$\sigma$ significance cutoffs. Each of the three probability density functions (blue lines plotted on the tops and side of the heat maps) are projections of the $\chi^2$ distributions. The solid red lines mark median values of each of the three parameters, while red dashed lines denote the upper and lower bounds of the 68.3\% (1-$\sigma$) credibility interval.}
    \label{fig:grid}
\end{figure}

\end{document}